\documentclass{moriond}

\def\be{\begin{equation}}
\def\ee{\end{equation}}
\def\bea{\begin{eqnarray}}
\def\eea{\end{eqnarray}}

\usepackage{color}
\usepackage{hyperref}
\definecolor{myblue}{RGB}{18, 124, 183}
\definecolor{mygreen}{RGB}{2, 158, 115}
\definecolor{myorange}{RGB}{222, 143, 5}
\definecolor{mydarkorange}{RGB}{211, 107, 21}
\definecolor{mypink}{RGB}{251, 175, 228}
\definecolor{mygray}{RGB}{161, 161, 161}

\newcommand{\gpn}[0]{$\gamma$-\textit{PhysNet}}

\newcommand{\Photo}{}

\begin{document}
\vspace*{4cm}
\title{Deep Learning for Cherenkov Astronomy: Performance of GammaLearn on LST-1}

\author{G.~Grolleron$^{*,a}$, C.~Plard$^a$, M.~Dell'aiera$^a$, T.~François$^a$, V.~Pollet$^a$, J.~Talpaert$^a$, S.~Caroff$^a$, T.~Vuillaume$^a$ \\ on behalf of the CTAO-LST Project}
\address{[a] Univ. Savoie Mont Blanc, CNRS, Laboratoire d'Annecy de Physique des Particules - IN2P3, 74000 Annecy, France}

\maketitle\abstracts{
The Cherenkov Telescope Array Observatory (CTAO) marks the next generation of Imaging Atmospheric Cherenkov Telescopes (IACTs), offering a sensitivity improvement of up to a factor of 10 over current instruments. Its first prototype, the Large-Sized Telescope (LST-1), is already in operation at the Roque de los Muchachos Observatory in La Palma, Spain. Deep learning methods have shown significant promise in reconstructing key properties of incident particles, such as energy, arrival direction, and type, using simulated data. Unlike traditional approaches that rely on simplified image shape parameters, deep learning can exploit the full temporal and charge information of the recorded events, providing enhanced performance, particularly at low energies ($\sim$20 GeV) accessible by LST-1. This capability is especially valuable for observing distant extragalactic sources like Active Galactic Nuclei, which are key to probing fundamental physics and cosmology. In this work, by producing sensitivity curves, we evaluate the performance of GammaLearn, a deep learning framework tailored for IACT data analysis, by comparing it to the standard analysis used with LST-1 and applying it to real observational data from LST-1.}

\section{Introduction}

The observation of very-high-energy (VHE) gamma-ray messengers has seen remarkable progress over the past few decades, driven by the success of ground-based Imaging Atmospheric Cherenkov Telescope (IACT) arrays. Building upon this legacy, the Cherenkov Telescope Array Observatory (CTAO) represents the next major leap in VHE astrophysics. CTAO will deliver a sensitivity improvement of a factor of five to ten depending of the energy. This will enable unprecedented investigations into non-thermal phenomena across the Universe. 
Operating at the La Palma site since 2019, the first Large-Sized Telescope (LST-1) has provided a crucial early glimpse into CTAO’s scientific potential. LST-1 has already demonstrated excellent performance, successfully detecting transients and challenging sources.

IACTs detect VHE gamma rays by capturing the Cherenkov light produced by extended air showers (EAS) in the atmosphere. The standard reconstruction pipeline for LST-1, \texttt{lstchain}~\citelow{lstchain_adass_2020}, relies on extracting Hillas parameters~\citelow{1985ICRC....3..445H} from calibrated camera images. Random Forest (RF) algorithms, trained on Monte Carlo (MC) simulations, are then applied to reconstruct the primary particle's nature, energy and direction. The robustness of this standard approach has been well established through extensive observations of the Crab Nebula~\citelow{2023ApJ...956...80A}.

To further push the limits of LST-1's reconstruction capabilities, recent studies have transitioned toward deep learning (DL) methodologies~\citelow{2021arXiv210804130V}. A major hurdle in applying DL to IACT data is the discrepancy between MC simulations and real observations, heavily influenced by the Night Sky Background (NSB) composed of the residual ambient light from the moon, stars, and atmospheric airglow. Its contamination depends on the field of view and the pointing. In a recent work~\citelow{Grolleron:2025Wb} we explored Unsupervised Domain Adaptation (UDA) techniques~\citelow{10.1145/3617233.3617279} and we proved that matching MC simulations to real data was still the best option. In this context, we utilize the \textsc{GammaLearn} framework~\citelow{Jacquemont:2019pw} to explore advanced Convolutional Neural Network (CNN) architectures. For each triggered event, the network ingests two integrated camera images over a 40 ns window: the integrated charge and the peak time per pixel. 

In a recent work~\citelow{Grolleron:2025Wb} we were constrained in the reconstruction by the limited generalization across all the sky. In this proceeding, we address a solution and we present a detailed evaluation of DL-based reconstruction for LST-1. In the first section, we provide a comprehensive study of the discrepancies between MC simulations and real data, and we derive an approach, using the generalizing capacity of CNN, to properly reconstruct observations of IACT. Before concluding, in the second section, we assess the performance of CNN (\gpn) architecture using LST-1 observations of the Crab Nebula as a standard reference.

\section{Solving MC and data discrepancies for a deep learning reconstruction\label{sec:mc_data}}

\begin{figure}[h]
    \centerline{\includegraphics[width=0.55\linewidth]{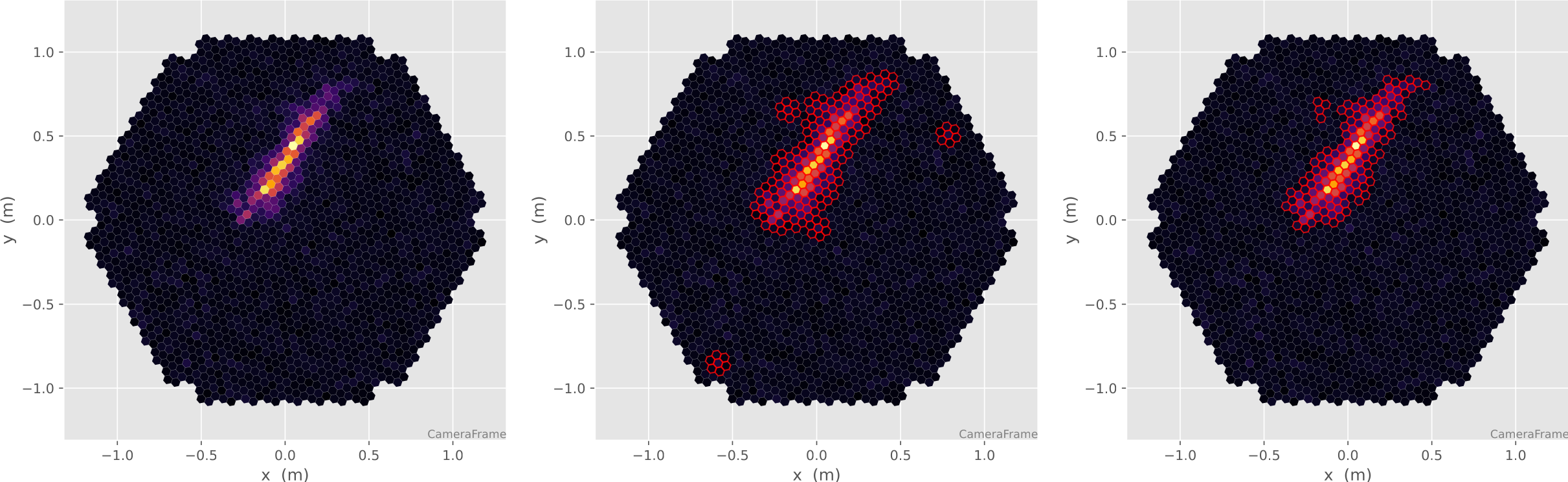}}
    \caption[]{\footnotesize Charge image of an LST-1 observed event shown at three levels of data reduction: all pixels (\textit{left}), DVR (\textit{center}), and \textit{Clean} (\textit{right}).}
    \label{fig:data_reduction}
\end{figure}

First, to reduce the amount of data for permanent storage, a simple data volume reduction  (DVR) scheme is applied, removing pixels which likely contain little or no Cherenkov light. Furthermore, image \textit{cleaning} techniques are typically coupled with RF algorithms to improve reconstruction robustness. We used the same \textit{cleaning} configuration as the one used in~\cite{2023ApJ...956...80A} and the DVR algorithm used here is a loose \textit{cleaning}. Figure~\ref{fig:data_reduction} illustrates the effect of DVR and \textit{cleaning} on a real data event; the number of retained pixels decreases drastically. Previous studies using MC simulations have shown that applying CNNs to DVR or \textit{cleaned} images significantly reduces reconstruction performance~\citelow{2025arXiv250207643F}. In this work, we compare the performance of \gpn~across these three levels of data reduction.

An extensive study was conducted to investigate the discrepancies between MC simulations of EAS and real data. The most effective approach is to closely match the background noise distribution in MC simulations to that of real data~\citelow{2021arXiv210804130V}. However, those discrepancies increase with the number of pixels retained after data reduction, suggesting that noise-dominated pixels are not accurately modeled in the simulations. These effect is critical because based on real data from the Crab Nebula, we demonstrated that accurately reproducing both the charge and time distributions in MC simulations is critical for CNN-based analyses~\citelow{Grolleron:2025Wb}.

Furthermore, two major differences persist between real data and MC simulations: the distributions are not centered on the same value, and their widths differ. As explained in a recent work~\citelow{Grolleron:2025Wb}, these effects occur because the absolute time depends on the data acquisition system and carries no intrinsic physical information. It would take a lot of work on the telescope trigger timing calibration side and the simulation side to perfectly match simulations to real data, for very little gain. The first issue can be easily resolved by using relative times. To achieve this, we explored two main techniques: centering the time of each event on either the charge-weighted time average or the time average of the three brightest pixels. The latter technique proved to be more robust. The issue regarding the distribution width was fixed simply by clipping the times between 4 and 25 ns. Several intervals have been tested and this one appeared to be the most robust. Ultimately, to resolve a smaller, second-order discrepancy and prevent \gpn~from focusing on MC specificity absent in real data, we applied a random Gaussian smearing to the event times during training, blurring the proton and gamma MC distributions.

\section{Performance of CNN evaluated with LST-1 observations of the Crab Nebula}

\begin{figure}
    \begin{minipage}{0.37\linewidth}
    \centerline{\includegraphics[width=\linewidth]{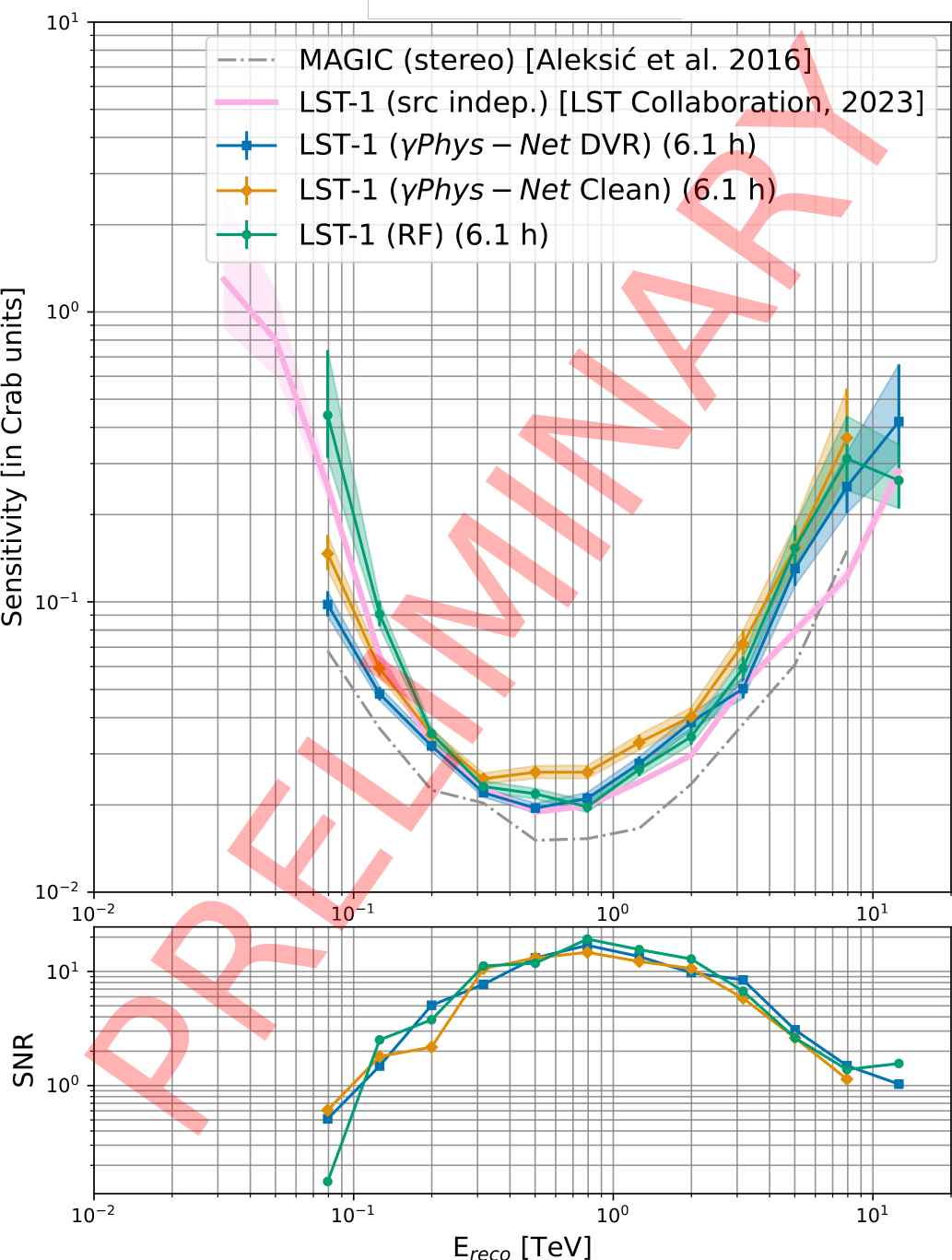}}
    \caption[]{\footnotesize Sensitivity curves for about 6h of observation, in percent of the Crab Nebula reference flux, for RF (\textcolor{mygreen}{green}), \gpn-\textit{clean} (\textcolor{myorange}{orange}), and \gpn-DVR (\textcolor{myblue}{blue}), compared with the sensitivity of MAGIC (\textcolor{mygray}{dashed-gray}) and of the LST-1 performance paper~\citelow{2023ApJ...956...80A} (\textcolor{mypink}{pink}). The errors take into account the systematics of the LST-1.}
    \label{fig:sensitivity}
    \end{minipage}
    \hfill
    \begin{minipage}{0.50\linewidth}
    \centerline{\includegraphics[width=\linewidth]{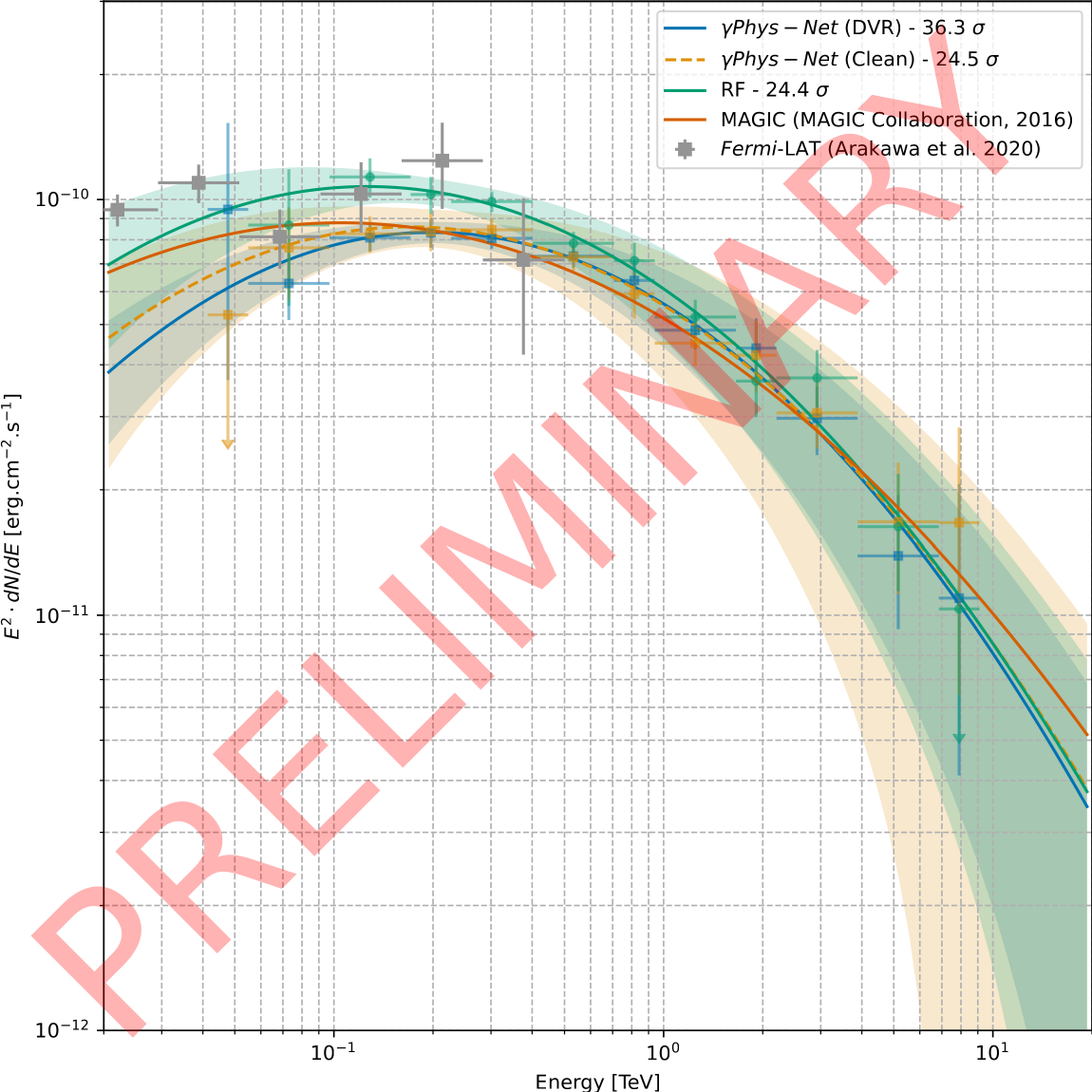}}
    \caption[]{\footnotesize Spectra and flux points of the Crab Nebula reconstructed with: RF (\textcolor{mygreen}{green}), \gpn-\textit{clean} (\textcolor{myorange}{orange}) and  \gpn-DVR (\textcolor{myblue}{blue}). They can be compared with \textit{Fermi}-LAT flux points (\textcolor{mygray}{gray}) and the spectrum published by the MAGIC collaboration (\textcolor{mydarkorange}{dark-orange}).}
    \label{fig:spectrum}
    \end{minipage}
\end{figure}

In a recent work~\citelow{Grolleron:2025Wb}, we trained \gpn~with MC simulations at a zenith angle of 20 degrees. The images were \textit{cleaned}, and the simulations were tuned with several NSB levels during training to help the model learn the NSB effects. The model demonstrated promising performance when applied to 50 hours of Crab Nebula observations with the LST-1. Here, we extended this approach: we used a large set of MC simulations covering all sky positions with zenith angles below 38 degrees. The transformation from absolute to relative time, as well as the clipping of the time images, have been crucial to achieving the performance presented in this article. To assess the \gpn~performance, about 6 hours of data taken between September 2025 to February 2026, with a zenith distance below 35 degrees are used to computed sensitivity curves (presented in Fig.~\ref{fig:sensitivity}). We used the algorithm from the LST-1 performance paper~\citelow{2023ApJ...956...80A} for RF, and for \gpn~trained on full, DVR, and \textit{cleaned} images (hereafter referred to as \gpn-full, \gpn-DVR, and \gpn-\textit{clean}, respectively). In the figure, the sensitivity evaluated for \gpn-full is not shown because its performance was comparable to the model trained with \textit{cleaned} images. This unexpectedly low performance is explained by the relatively limited match between MC simulations and real data for the noise-dominated pixels, which constitute the majority of the pixels. Regarding the other models, the results can be summarized as follows:
\begin{itemize}
    \item[] \gpn-\textit{clean} performs better than RF for energies below 200 GeV but worse at higher energies, 
    \item[] \gpn-DVR outperforms both RF and \gpn-\textit{clean} below 1 TeV, reaching a factor of 3 improvement at 80 GeV, and is compatible with RF at energies above 1 TeV.
\end{itemize} 
Moreover, as shown in the lower panel of Fig.~\ref{fig:sensitivity}, the signal-to-noise ratio is better at low energies for \gpn. This is extremely promising because it could help reduce the systematic errors that dominate at low energies. 

The Crab Nebula spectrum was estimated using cuts on discriminating variables, which were evaluated on MC simulations with a selection of gamma-like events of 70\% and a 70\% containment radius, as presented in Fig.~\ref{fig:spectrum}. We can see that, within statistical errors, the three methods are compatible; however, the advantage of \gpn-DVR is clearly visible at low energies, where a flux point is successfully computed at 50 GeV (indicating at least a 3$\sigma$ detection). Furthermore, the overall spectrum is more accurate, even above 1 TeV. Regarding the \gpn-\textit{clean} model, a reduction in the accuracy of the spectrum estimation is visible. Nevertheless, this model remains an interesting alternative to RF below a few hundred GeV. 

\section{Conclusion}

In this article, we presented the first mature \gpn~reconstruction applied to LST-1 data of the Crab Nebula. We proved that this architecture is able to generalize properly across MC simulations under various zenith angles and with several levels of NSB. We compared the \gpn~performance across several levels of data reduction and obtained the best results with the DVR format requested by CTAO. This performance is reduced when using \textit{cleaned} images. Furthermore, by using DVR images, we achieved significantly better sensitivity as well as a more accurate spectrum compared to the standard RF reconstruction. Moving forward, we need to validate these results with a larger data sample and with other sources observed by the LST-1.

\tiny
\section*{Acknowledgments}

This project is supported by the facilities offered by the Univ. Savoie Mont Blanc - CNRS/IN2P3 MUST computing center
This project was granted access to the HPC resources of IDRIS under the allocation 2020-AD011011577 made by GENCI
This project is supported by the computing and data processing ressources from the CNRS/IN2P3 Computing Center (Lyon - France)
We gratefully acknowledge the support of the NVIDIA Corporation with the donation of one NVIDIA P6000 GPU for this research.
We gratefully acknowledge financial support from the agencies and organizations listed here: \href{https://www.ctao.org/for-scientists/library/acknowledgments/}{https://www.ctao.org/for-scientists/library/acknowledgments/}.
This project has received funding from the Fondation Université Savoie Mont-Blanc
This project has received funding from the European Union’s Horizon 2020 research and innovation programme under grant agreement No 653477
This project has received funding from the European Union’s Horizon 2020 research and innovation programme under grant agreement No 824064
This project has received funding from the European Union Horizon Programme call INFRAEOSC-03-2020 - Grant Agreement No 101017536
The author acknowledges the support of the French Agence Nationale de la Recherche (ANR) under reference ANR-23-CE31-0021 (DIRECTA)

\normalsize
\section*{References}
\bibliography{moriond}


\end{document}